\documentclass[aps, 10pt, prd,
              notitlepage, twocolumn, superscriptaddress,
              nofootinbib,floatfix]{revtex4-1}

\usepackage{amsmath}
\usepackage{aas_macros}
\usepackage{amssymb}
\usepackage{amsfonts}
\usepackage[utf8]{inputenc}
\usepackage[T1]{fontenc}
\usepackage{mathrsfs}
\usepackage{anyfontsize}

\usepackage[linktocpage,breaklinks]{hyperref}
\usepackage[usenames,dvipsnames]{xcolor}

\usepackage{tensor}
\usepackage{bm}

\usepackage{graphicx}
\usepackage{epsfig}
\usepackage{epstopdf}

\usepackage{natbib}
\usepackage{hyperref}

\hypersetup{colorlinks=true,
            citecolor=black,
            linkcolor=black,
            urlcolor=black}

\newcommand{\mg}{\mathfrak{M}}

\newcommand{\rc}{r_{\ast}}
\newcommand{\dd}{\textrm{d}}
\newcommand{\bg}{\textbf{g}}

\newcommand{\calp}{\mathscr P}

\newcommand{\calm}{\mathscr M}
\newcommand{\calM}{\mathcal{M}}
\newcommand{\ptovt}{post-TOV }
\newcommand{\msun}{$_{\odot}$}
\newcommand{\ppeb}{\mathfrak{b}}

\begin{document}

\title{More than the sum of its parts: \\
       combining parameterized tests of extreme gravity}

\begin{abstract}
We connect two formalisms that describe deformations away from general
relativity, one valid in the strong-field regime of neutrons stars and
another valid in the radiative regime of gravitational waves: the
post-Tolman-Oppenheimer-Volkoff and the parametrized-post-Einsteinian
formalisms respectively.
We find that post-Tolman-Oppenheimer-Volkoff deformations of the exterior
metric of an isolated neutron star induce deformations in the orbital
binding energy of a neutron star binary.
Such a modification to the binding energy then percolates into the
gravitational waves emitted by such a binary, with the leading-order
post-Tolman-Oppenheimer-Volkoff modifications introducing a second
post-Newtonian order correction to the gravitational wave phase.
The lack of support in gravitational wave data for general relativity
deformations at this post-Newtonian order can then be used to place
constraints on the post-Tolman-Oppenheimer-Volkoff parameters.
As an application, we use the binary neutron star merger event GW170817 to
place the constraint $-2.4 \leq \chi \leq 44$ (at 90\% credibility) on a
combination of post-Tolman-Oppenheimer-Volkoff parameters.
We also explore the implications of this result to the possible
deformations of the mass-radius relation of neutron stars allowed within
this formalism.
This work opens the path towards theory-independent tests of gravity,
combining astronomical observations of neutron stars and gravitational wave
observations.
\end{abstract}

\author{Hector O. Silva}
\email{hector.okadadasilva@montana.edu}
\affiliation{eXtreme Gravity Institute,
Department of Physics, Montana State University, Bozeman, Montana 59717, USA}

\author{Nicol\'as Yunes}
\email{nicolas.yunes@montana.edu}
\affiliation{eXtreme Gravity Institute,
Department of Physics, Montana State University, Bozeman, Montana 59717, USA}

\date{{\today}}

\maketitle

\section{Introduction}
\label{sec:intro}

Neutron stars are one of the prime objects in nature for confronting our
understanding of fundamental physical interactions against
observations~\cite{Watts:2016uzu,Berti:2015itd,Doneva:2017jop}.
Their small size (radius around $\approx 12$ km) and large mass ($\approx 1.4$
M$_\odot$) result in densities at their core that can exceed that of nuclear
saturation density, at which hadronic matter can transmute into exotic
forms, by 10 orders of magnitude~\cite{Lattimer:2006xb}.
Neutron stars are also extreme gravity objects, second only to black holes in
the strength of their gravitational potential and spacetime curvature, with
fields that exceed those that we experience in the neighborhood of our Solar
System by 9 orders of magnitude.
The strong-field regime of neutron stars, critical in determining their
structure and
stability~\cite{Oppenheimer:1939ne,Tolman:1939jz,Bonolis:2017}, demands the
use of relativistic gravity to describe these stars, with Einstein's general
relativity (GR) as our canonical theory for doing so.
Moreover, neutron stars unlike black holes, allow us to probe how matter couples
with the very fabric of spacetime in the strong-field
regime~\cite{Delsate:2012ky}.

The piercing power of neutron stars as tools to test our understanding of
nature is amplified when they are found in binaries. From the discovery of
the very first binary pulsar~\cite{Hulse:1974eb} and the confirmation that
its orbital period decays in agreement with GR predictions, through the
emission of gravitational waves~\cite{Damour:2014tpa}, to the spectacular
detection of the first binary neutron star merger event
GW170817~\cite{TheLIGOScientific:2017qsa} by the LIGO/Virgo collaboration
(LVC), neutron star binaries have been in the forefront of experimental
gravity in astronomical settings with implications to cosmology
included~\cite{Yunes:2013dva,Sakstein:2017xjx,Baker:2017hug,Ezquiaga:2017ekz,Creminelli:2017sry}

Experimental tests of relativistic gravity have a long
history~\cite{Will:1993ns,Will:2014kxa} and can basically be carried out in
two ways.
In the first approach, one assumes a particular theory, whose predictions
are worked out and then tested against observations.
In the second approach, one introduces deformations to the predictions or
solutions of GR, in a particular regime of the theory, and one then works
out the observational consequences of these deformations to confront them
against observations.
Both approaches have been successful in aiding our understanding of the
nature of gravity.
An example of the first approach is the ruling out of Nordstr\"om's theory
of gravity (a predecessor to GR), which for example fails to predict the
deflection of light by the Sun~\cite{Dyson:1920cwa,1974Mehra,Will:2014zpa}.
An example of the second approach is the parametrized post-Newtonian
framework (ppN)~\cite{Will:1971zzb,Will:1972zz,Nordtvedt:1972zz}, which
allowed us to test GR against a myriad of new Solar System tests starting
in the 1960s, although early ideas date back to
Eddington~\cite{Eddington:Book}.

Can we combine parametrized tests of gravity that involve observations of the strong-field
gravity created by isolated neutron stars with those that involve the radiative and dynamical fields
generated in the coalescence of neutron star binaries?
The purpose of this paper is to build a bridge between two parametrizations
for tests of GR: the parametrized post-Tolman-Oppenheimer-Volkoff (post-TOV)
formalism~\cite{Glampedakis:2015sua,Glampedakis:2016pes} (which
parametrizes deviations to the stellar structure of isolated neutrons stars) and the
parametrized-post-Einsteinian (ppE)
formalism~\cite{Yunes:2009ke,Yunes:2016jcc} (which parametrizes deviations
to GR in the inspiral, merger and ringdown of compact binary
coalescence).
This bridge provides a theory-independent framework to combine constraints on
deviations to GR from the observation of the bulk properties of neutron
stars and from the generation and propagation of gravitational
waves produced in the coalescence of binary neutron stars.

\begin{figure}[t]
    \includegraphics[width=\columnwidth]{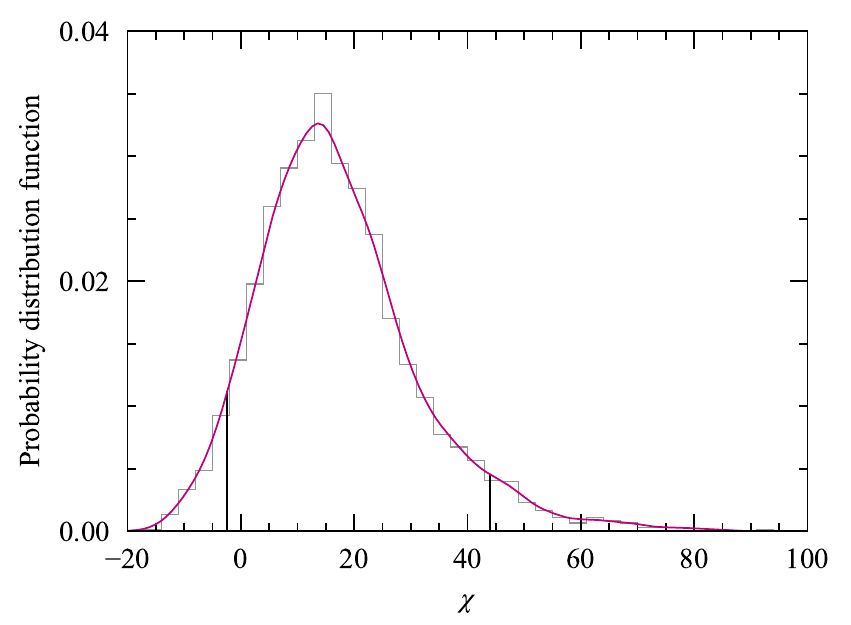}
\caption{Marginalized posterior distribution for the $\chi$ post-TOV parameter,
obtained from the Markov-chain-Monte-Carlo (MCMC) samples released by the
LVC for the GW170817  event.
The 90\% credible interval bound on $\chi$ is $-2.4 \leq \chi \leq 44$, as indicated
by the vertical lines.
The lower support at zero is not an evidence of a deviation from GR as
explained in the text, but rather it reflects a similarly skewed posterior distribution for
$\delta \phi_4$, which peaks away from zero due to degeneracies between the
various binary parameters and nonstationarity of the detector noise.
The long tail of the distribution is produced by a similar tail in the marginalized
posterior for $\delta\phi_{4}$, the parameter that encodes deformations in the
gravitational wave Fourier phase at 2PN order (see Fig. 1 in
Ref.~\cite{Abbott:2018lct}).}
\label{fig:chi_posterior}
\end{figure}

The connection between both formalisms is only possible by realizing that
the modified exterior spacetime of neutron stars in the post-TOV formalisms
affects the binding energy of a neutron star
binary~\cite{Glampedakis:2016pes}, and thus, the gravitational waves that
such a binary emits~\cite{Yunes:2009ke}.
This modification to the binding energy or the gravitational waves emitted
can be mapped onto the ppE framework, which we have extended here to
encompass a wider set of modifications to the conservative sector of the
binary's Hamiltonian.
This allows a particular combination of post-TOV parameters $\chi$ [defined
in Eq.~\eqref{eq:chi_def}] to be mapped to the ppE modification to the
gravitational wave Fourier phase $\delta \psi_{\rm ppE}$
[cf.~Eqs.~\eqref{eq:ppe_phase}
and~\eqref{eq:ppe_beta_and_b_final_mapping}].
We find that $\chi$ modifies the gravitational wave evolution at second
post-Newtonian order (2PN)\footnote{The PN formalism is one in which the
field equations are solved perturbatively as an expansion in weak
fields and small velocities. A term of $N$PN order is of
${\cal{O}}(v^{2N}/c^{2N})$ relative to the leading-order term, with $v$
the orbital speed and $c$ the speed of light~\cite{Blanchet:2013haa}}.

The lack of support in gravitational wave data for a GR deformation then
allows for constraints on deformations of the exterior metric of isolated
neutron stars. In particular, the constraints on GR modifications obtained
by the LVC~\cite{Abbott:2018lct} for the binary-neutron star gravitational
wave event GW170817~\cite{LIGOScientific:2018mvr} can be used to place the
first observational constraint on $\chi$, namely $-2.4 \leq \chi \leq 44$
at 90\% credibility (see Fig.~\ref{fig:chi_posterior}).
This result strengthens the case for compact binary mergers as laboratories
to test GR, something which would otherwise be very hard (if not
impossible) with \textit{only} mass and radius measurements of isolated
neutron stars due to strong degeneracies between matter and strong-field
gravity. We provide explicit examples of this degeneracy by computing the
post-TOV deformations to the mass-radius curves within $-2.4 \leq \chi \leq
44$ for a fixed equation of state.

The remainder of the paper presents the details that led to the results
summarized above and it is organized as follows.
In Sec.~\ref{sec:intro} we briefly overview the post-TOV and ppE
formalisms, establishing the connection between the two.
Next, in Sec.~\ref{sec:constraints} we use the public data on tests of GR
with GW170817 released by LVC to place constraints on a combination of
post-TOV parameters.
In Sec.~\ref{sec:mr} we discuss the allowed deformation on the mass-radius
curves of neutron stars under this constraint, discussing in detail the
degeneracies between matter and strong gravity.
In Sec.~\ref{sec:conclusions}, we present our conclusions and outline some
directions for which our work can be extended. Throughout this work we use
geometric units $G = 1 = c$ and use a mostly plus metric signature.

\section{From Post-TOV to ppE}

Let us start by briefly reviewing the \ptovt formalism developed in
Refs.~\cite{Glampedakis:2015sua,Glampedakis:2016pes} and the ppE formalism
introduced in Ref.~\cite{Yunes:2009ke} and expanded
in~\cite{Chatziioannou:2012rf}.

\subsection{Overview of the post-TOV formalism}
\label{sec:ptov}

The idea behind the post-TOV formalism is quite simple. The formalism is based
on the observation that the structure of static, spherically symmetric stars in
GR is determined by only two differential equations:
\begin{subequations}
\begin{align}
    \left(\frac{\dd p}{\dd r}\right)_{\rm GR} &= - \frac{(\epsilon + p)}{r^2}\,
    \frac{(m+4\pi r^3 p)}{1 - 2 m / r}\,,
    \\
    \left(\frac{\dd m}{\dd r}\right)_{\rm GR} &= 4 \pi r^2 \epsilon\,,
\end{align}
\end{subequations}
which respectively govern the pressure and mass gradients within the star.
Here, $r$ is the circumferential radius, $m$ the mass function, $p$ the pressure
and $\epsilon$ the total energy density. The latter two variables are assumed to
be related through a barotropic equation of state (EOS), i.e. $p = p(\epsilon)$.
For later convenience we recall that $\varepsilon$ can be written as
$\varepsilon = \rho(1 + \Pi)$, where $\rho$ is the baryonic rest-mass
density and $\Pi$ the internal energy per unit baryonic mass.

The post-TOV formalism augments these equations to the form
\begin{subequations}
\begin{align}
    \frac{\dd p}{\dd r} &= \left(\frac{\dd p}{\dd r}\right)_{\rm GR}
    - \frac{\rho m}{r^2}\left( \calp_{1} + \calp_{2} \right)\,
    \\
    \frac{\dd m}{\dd r} &= \left(\frac{\dd m}{\dd r}\right)_{\rm GR}
    + 4 \pi r^2 \rho \left( \calm_1 + \calm_2 \right)\,
\end{align}
\label{eq:postTOV}
\end{subequations}
where the first set of post-TOV corrections is
\begin{subequations}
\begin{align}
    \calp_1 &\equiv \delta_{1} \frac{m}{r} + 4 \pi \delta_2 \frac{p r^3}{m} \,,
    \\
    \calm_1 &\equiv \delta_{3} \frac{m}{r} + \delta_{4} \Pi\,,
\end{align}
\end{subequations}
and the second set is
\begin{subequations}
\begin{align}
    \calp_2 &\equiv \pi_1 \frac{m^2}{\rho r^5} + \pi_2 \frac{m^2}{r^2}
    + \pi_3 p r^2 + \pi_4 \frac{\Pi p}{\rho}\,,
    \\
    \calm_2 &\equiv \mu_1 \frac{m^2}{\rho r^5} + \mu_2 \frac{m^2}{r^2}
    + \mu_3 p r^2 + \mu_4 \frac{\Pi p}{\rho} + \mu_5 \Pi^{3} \frac{r}{m} \,,
    \nonumber \\
\end{align}
\label{eq:second_set}
\end{subequations}
where $\delta_i$, $\pi_i$ and $\mu_i$ are all dimensionless
constants.

The first set ($\calp_1$, $\calm_1$) arises from the ppN stellar structure
equations~\cite{Wagoner1974,Shapiro1976,Ciufolini1983,Glampedakis:2015sua}.
These non-GR terms in the post-Newtonian regime were then added to the full GR
equations to capture effects of modifications to GR.
Indeed, the parameters $\delta_{i}$ are all related to the usual
ppN parameters via $\delta_1 \equiv 3(1+\gamma)-6\beta+\zeta_2$,
$\delta_2 \equiv \gamma - 1 + \zeta_4$,
$\delta_3 \equiv - (1/2)(11+\gamma-12\beta+\zeta_2-2\zeta_4)$ and $\delta_4 \equiv \zeta_3$.
Solar System constraints impose $|\delta_i| \ll 1$, yielding $\calp_1 \ll 1$ and $\calm_{1} \ll
1$ in Eq.~\eqref{eq:postTOV}, and thus, we will here only study
the second set of post-TOV corrections.

The second set ($\calp_2$, $\calm_2$) represents 2PN corrections which can be
written in terms of fluid and metric variables. As explained in detail in
Ref.~\cite{Glampedakis:2015sua}, the 2PN terms which can be constructed from
these primitive quantities can be gathered in five ``families,'' each with an
infinite number of terms and with each family yielding a distinctive change to
the mass-radius relation of neutron stars.
Fortunately, 2PN terms belonging to each family exhibit qualitatively the same
radial profiles inside a star. This translates into terms belonging to the same
family affecting the mass-radius relations in a self-similar manner
(cf.~\cite{Glampedakis:2015sua}, Figs. 3, 6 and 7).
This fact allows one to choose \textit{a single representative} member from each
family to be included to the TOV equations.
The criteria used in~\cite{Glampedakis:2015sua} to make this choice was that
of overall magnitude of the modification (relative to other terms in the same family) and
simplicity of the analytic form of the term.

Equation~\eqref{eq:postTOV} is sufficient to determine the \textit{interior}
of the star and its bulk properties i.e. the (Schwarzschild) enclosed mass $\mg$
$[\equiv m(R)]$ and the radius $R$ [location $r=R$ at which $p(R) = 0$
when integrating the post-TOV equation outwards from $r=0$.].
In~\cite{Glampedakis:2016pes}, the \textit{exterior} problem was addressed and
it was found that the post-TOV equations result in a
\textit{post-Schwarzschild} exterior metric given by
\begin{subequations}
\begin{align}
    g_{tt} &= - \left(1 - \frac{2M}{r}\right) + \frac{2\chi}{3}  \left( \frac{M}{r}\right)^3\,,
    \\
    g_{rr} &= \left(1 - \frac{2M}{r}\right)^{-1} - 4\pi\mu_1 \left( \frac{M}{r}\right)^3\,,
\end{align}
\label{eq:ptov_exterior}
\end{subequations}
where
\begin{equation}
\chi \equiv \pi_2 - \mu_2 - 2 \pi \mu_1\,,
\label{eq:chi_def}
\end{equation}
is a combination of the \ptovt parameters and
\begin{equation}
    M = \mg\left[1 + 2 \pi \mu_1 \left(\frac{\mg}{R}\right)^2\right]\,,
    \label{eq:adm_mass}
\end{equation}
is the Arnowitz-Misner-Deser mass of the star.
Equation~\eqref{eq:adm_mass} was obtained under the restriction that $\mu_{1} \in
[-1.0, 0.1]$, outside of which the calculation of $M$ requires solving a
transcendental equation and for which the exterior metric cannot be written
analytically in the simpler form~\eqref{eq:ptov_exterior}.

The fact that $M \neq \mg$ is not unusual in modified theories of gravity
(see e.g.~\cite{Damour:1993hw}).  In theories beyond GR, contributions to
the star's mass due to the presence of new degrees of freedom, such as
scalar or vector fields arise, although this is not always the
case~\cite{Cisterna:2015yla,Maselli:2016gxk,Cisterna:2016vdx}. We stress that it is $M$, not $\mg$, which
would be observationally inferred, e.g. by using Kepler's law.

In dynamical situations, such as in the motion of a neutron star binary,
these additional degrees of freedom can be excited, and thus, they can open new
radiative channels for the system to lose energy, modifying the binary's
dynamic. As formulated, the post-TOV formalism cannot account for the
presence of extra fields and hence the radiative loses of the binary will
be the same as in GR. On the other hand, since the exterior spacetime is
different from that of Schwarzschild, the conservative sector of the binary
motion will be different.

As we will see next, the ppE formalism aims to capture generic deviations
from GR to both sectors. This will allow us to obtain a mapping between the
parameters (that control these deviations) in both formalisms.

\subsection{Overwiew of the ppE formalism}
\label{sec:ppe}

The ppE formalism was developed to capture generic deviations from GR in the
gravitational waves emitted by a binary system~\cite{Yunes:2009ke}. These
deviations can be separated into those that affect the conservative sector
(e.g.~the binding energy of the orbit) and the dissipative sector (e.g.~the flux
of energy). In previous work, the conservative sector was modified in a rather
cavalier way, making some assumptions about the structure of the deformations.
Let us then here relax some of these assumptions and rederive the modifications.

We begin with the Hamiltonian for a two-body system in the center of mass frame,
working to leading order in the post-Newtonian approximation  and to leading order
in the GR deformation:
\begin{align}
\label{eq:Hamiltonian}
H &= p_{\alpha}\, p^{\alpha}
  \nonumber \\
  &=
\frac{p_{r}^{2}}{2 \mu} \left(1 + \delta p_{r}\right)
+ \frac{p_{\phi}^{2}}{2 \mu r^{2}} \left(1 + \delta p_{\phi}\right)
- \frac{\mu m}{r} \left(1 + \delta U\right)\,,
\end{align}
where
$r$ is the relative separation of the binary, $\mu = m_{1} m_{2}/m$ is the
reduced mass, with $m_{1,2}$ the component masses and $m=m_{1}+m_{2}$ the
total mass, and $p_{r}$ and $p_{\phi}$ are the generalized momenta
conjugate to the radial and azimuthal coordinates.

The functions $(\delta U, \delta p_{r}, \delta p_{\phi})$  characterize the
deformation to the standard Newtonian Hamiltonian. For the purposes of this
work, we will parametrize these deformations as
\begin{align}
\label{eq:ansatz}
\delta U &= A \left(\frac{m}{r}\right)^{a}\,,
\quad
\delta p_{r} = B \left(\frac{m}{r}\right)^{b}\,,
\quad
\delta p_{\phi} = C \left(\frac{m}{r}\right)^{c}\,,
\end{align}
where $(A,B,C)$ control the magnitude of the deformation (assumed small
here), while $(a,b,c)$ control the character of the deformation. We will
also here assume that $a = b = c$, meaning that all deformations enter at
the same post-Newtonian order, and we will discuss later how to relax this
assumption.  Physically, we can think of $(\delta U, \delta p_{r}, \delta
p_{\phi})$ as modifying the $(t,t)$, $(r,r)$ and $(\phi,\phi)$ components
of the metric respectively. Notice also that if $\delta p_{\phi} \neq 0$,
then the radius $r$ and the angle $\phi$ are not your usual circumferential
radius and azimuthal angle (though they are related to them via a
coordinate transformation).

With this at hand, we can now derive the constants of the motion and the
field equations.  Assuming the Hamilton equations hold, there are two
constants of the motion associated with time translation and
azimuthal-angle translation invariance. The former is simply the
Hamiltonian itself, which for a binary is the binding energy $E_{\rm b}$.
The latter is the angular momentum of the orbit, which we can define as $L
\equiv p_{\phi}/\mu$. The azimuthal component of the generalized momenta
can be obtained from
\begin{equation}
\dot{\phi} = \frac{\partial H}{\partial p_{\phi}} = \frac{p_{\phi}}{\mu \, r^{2}} \left(1 + \delta p_{\phi}\right)+ \frac{p_{\phi}^{2}}{2 \mu \, r^{2}}  \frac{\partial \delta p_{\phi}}{\partial p_{\phi}} \,,
\end{equation}
which then leads to
\begin{equation}
L = \omega \, r^{2} \left(1 - \delta p_{\phi}\right)\,,
\end{equation}
where have used the definition $\omega \equiv \dot{\phi}$, and because
$\delta p_{\phi}$ was assumed to be independent of $p_{\phi}$ by
Eq.~\eqref{eq:ansatz}.

With this at hand, we can now derive the radial equation of motion in
reduced order form.  We begin by evaluating $\dot{r}$, which by Hamilton's
equation is simply $(p_{r}/\mu) (1 + \delta p_{r})$, where again we have
used that $\delta p _{r}$ was assumed to be independent of $p_{r}$ from
Eq.~\eqref{eq:ansatz}.  We can then rewrite Eq.~\eqref{eq:Hamiltonian} as
\begin{align}
\label{eq:Hamiltonian2}
 \frac{\dot{r}^{2}}{2} \left(1 - \delta p_{r}\right) =
 \frac{E_{\rm b}}{\mu} +  \frac{m}{r} \left(1 + \delta U\right) - \frac{L^{2}}{2 r^{2}} \left(1 + \delta p_{\phi}\right) \equiv V_{\rm eff}\,.
\nonumber \\
\end{align}
Note that $\delta p_{r}$, which is associated with a deformation of the
$(r,r)$-component of the metric does not affect the location in phase space
where $\dot{r} = 0$ (or equivalently where $V_{\rm eff} = 0$).

Before we can find what the binding energy of the orbit is as a function of
the orbital angular frequency, we must determine what the energy and the
angular momentum of a circular orbit in this perturbed spacetime is.  We
can do so by setting $V_{\rm eff} = 0$ and $\dd V_{\rm eff}/ \dd r = 0$ and
solving for $E_{\rm b}$ and $L$, which yields
\begin{align}
\label{eq:E-pre}
\frac{E_{\rm b}}{\mu} &= - \frac{m}{2 r} \left[1 + A \left(1 - a\right) \left(\frac{m}{r}\right)^{a} + C \frac{c}{2} \left(\frac{m}{r}\right)^{c} \right]\,,
\\
\label{eq:L-pre}
L &= \sqrt{m r} \left[1 + \frac{A}{2} \left(1 + a\right) \left(\frac{m}{r}\right)^{a} - \frac{C}{2} \left(1 + \frac{c}{2}\right) \left(\frac{m}{r}\right)^{c} \right]\,.
\nonumber \\
\end{align}
From the above expression for $L$, we can solve for $\omega(r)$ as well as
$r(\omega)$ (i.e.~the modification to Kepler's third law) to find
\begin{align}
\frac{m}{r} &= \left(m \omega\right)^{2/3}  \left[1 - \frac{A}{3} \left(1 + a\right) \left(m \omega\right)^{2a/3}
\right. \nonumber \\
&\quad - \left. \frac{C}{3} \left(1 - \frac{c}{2}\right) \left(m \omega\right)^{2c/3} \right]\,.
\label{eq:mod-kep}
\end{align}
Using this in Eq.~\eqref{eq:E-pre}, we then find the final expression
\begin{align}
\frac{E_{\rm b}}{\mu} &= - \frac{1}{2} \left(m \omega\right)^{2/3} \left[1 + \frac{2A}{3} \left(1-2a \right) \left(m \omega\right)^{2a/3}
\right. \nonumber \\
&\quad - \left.
    \frac{C}{3} \left(1 - 2 c \right)  \left(m \omega\right)^{2c/3} \right]\,,
\label{eq:bindng-e-of-w}
\end{align}

Reference~\cite{Chatziioannou:2012rf} carried out a similar calculation,
except that in their calculation, the whole Newtonian effective potential
was modified by the same term, namely
\begin{equation}
\label{eq:effpot-25}
V_{\rm eff}^{\mbox{\tiny \cite{Chatziioannou:2012rf}}} = \left(- \frac{m}{r} + \frac{L^{2}}{2 r^{2}} \right) \left[ 1 + A^{\mbox{\tiny \cite{Chatziioannou:2012rf}}} \left(\frac{m}{r}\right)^{p} \right]\,.
\end{equation}
Such a modification lead to a binding energy of the form~\cite{Chatziioannou:2012rf}
\begin{align}
    \frac{E_{\rm b}^{\mbox{\tiny \cite{Chatziioannou:2012rf}}}}{\mu} &= -\frac{1}{2} \left(m \omega\right)^{2/3} \left[1 - \frac{1}{3} A^{\mbox{\tiny \cite{Chatziioannou:2012rf}}}(2p-3) \left(m \omega\right)^{2p/3}\right]\,.
    \label{eq:ppe_binding}
\end{align}
From this, Ref.~\cite{Chatziioannou:2012rf} showed that the gravitational
waves emitted by a binary, assuming the dissipative sector is not modified
(i.e~the flux of energy is the same as that in GR), and assuming
gravitational waves contain the same two polarizations as in GR, lead to a
Fourier detector response (in the stationary phase approximation) of the
form
\begin{equation}
\tilde{h} = {\cal{A}}(f) e^{i \Psi(f)}\,,
\end{equation}
where ${\cal{A}}$ is the Fourier amplitude and $\Psi$ is the Fourier phase. The
latter can be decomposed into $\Psi = \Psi_{\rm GR} + \delta \psi$, where
$\Psi_{\rm GR}$ is the Fourier phase in GR, while the GR deformation is
\begin{equation}
    \delta \psi = \frac{5}{32}\, A^{\mbox{\tiny \cite{Chatziioannou:2012rf}}}  \,
    \frac{(2p^2-2p-3)}{(4-p)(5-2p)} \eta^{-2p/5} u^{2p-5}\,,
\end{equation}
where
\begin{equation}
    u= (  \pi \calM f )^{1/3}\,,
\end{equation}
and $f$ is the gravitational wave frequency.

Given the similarities in the calculations, the easiest way forward is to map the
results of Ref.~\cite{Chatziioannou:2012rf} to the modifications we are considering here.
Comparing the binding energies in Eqs.~\eqref{eq:ppe_binding} and~\eqref{eq:bindng-e-of-w}, we see that
\begin{align}
A^{\mbox{\tiny \cite{Chatziioannou:2012rf}}}
=
2A\, \frac{1-2a }{3-2a}
-
C\, \frac{1-2a}{3-2a}\,,
\end{align}
and where we have used that $a = c = p$. We then see clearly that the
change in the Fourier phase is
\begin{align}
    \delta \psi &= \frac{5}{32}\, \left( 2A\, \frac{1-2a }{3-2a}
                - C\, \frac{1-2a}{3-2a} \right)
                \nonumber \\
                &\quad \times \frac{(2a^2-2a-3)}{(4-a)(5-2a)} \,
                \eta^{-2a/5} u^{2a-5}\,.
\end{align}

This deformation arising from a GR correction to the binding energy can be
mapped to the ppE waveform as follows. Noting that the ppE phase
is~\cite{Yunes:2016jcc}
\begin{equation}
    \delta \psi_{\rm ppE} = \beta (\pi \calM f)^{\mathfrak{b}/3}\,,
    \label{eq:ppe_phase}
\end{equation}
we then realize that
\begin{subequations}
\begin{align}
\beta &=  \frac{5}{32}\, \left( 2A \,\frac{1-2a }{3-2a}
- C\, \frac{1-2a}{3-2a} \right)
\nonumber \\
&\quad \times
\frac{(2a^2-2a-3)}{(4-a)(5-2a)}\, \eta^{-2a/5}\,,
\\
\ppeb &= 2 a - 5\,.
\end{align}
\label{eq:ppe_beta_and_b}
\end{subequations}
Therefore, a ppE constraint on $\beta$ for a given value of $\ppeb$ given a
gravitational wave observation that is consistent with GR can be
straightforwardly mapped to a constraint on $A$ given a value of $a$.

\subsection{Relating the parameters in both formalisms}

Several paths are possible to relate the post-TOV and the ppE formalisms.
The path we choose here is to compare the binding energy and angular
momentum of a binary system composed of neutron stars whose metrics in
isolation would take the form of Eq.~\eqref{eq:ptov_exterior}. This can be
achieved by transforming from the two-body problem to an effective one-body
problem, in which a test particle of mass $\mu = m_{1} m_{2}/m$ moves in a
background of mass $m = m_{1} + m_{2}$. Let us then consider the geodesic
motion of a test particle in a generic (but still stationary and
spherically symmetric) background.

Consider the line element
\begin{equation}
    \dd s^2 = - f(r) \dd t^2 + h(r) \dd r^2 + r^2 (\dd\theta^2 +
    \sin^2\theta\, \dd\phi^2)\,,
\end{equation}
where the metric functions $f$ and $h$ are decomposed as $f(r) = f_0(r) +
\varepsilon f_1(r)$ and $h(r) = f^{-1}_0(r) + \varepsilon h_1(r)$, and
where $\varepsilon$ is a small bookkeeping parameter. In
Appendix~\ref{app:binding} we present a detailed analysis of geodesic
circular motion in such a perturbed metric, and we compute the change to
the binding energy $E$ and the angular momentum $L$ of the orbit.
Identifying $f_0 = 1 - 2M/r$, $f_1= - (2\chi/3) (M/r)^3$, substituting
these expressions into Eqs.~\eqref{eq:e_final} and~\eqref{eq:l_final}, and
expanding both in $\varepsilon \ll 1$ and in $M/r \ll 1$, we find
\begin{align}
    \frac{E_{\rm b}}{\mu} &\equiv \frac{E - 1}{\mu}
              = -\frac{m}{2 r}\left(1 - \frac{1}{3}\,\chi\,\frac{m^2}{r^2}\right)\,,
    \label{eq:e_b_intermediate}
    \\
     \frac{L}{\mu} &= \sqrt{m r} \left( 1 + \frac{1}{2}\, \chi\, \frac{m^2}{r^2} \right)\,,
     \label{eq:L_intermediate}
\end{align}
where $\chi$ is a post-TOV parameter.

We can now compare Eq.~\eqref{eq:e_b_intermediate} to Eq.~\eqref{eq:E-pre}
and Eq.~\eqref{eq:L_intermediate} to~\eqref{eq:L-pre} to find what $A$, $C$
and $a$ are in the post-TOV formalism.  Doing so, we find that $A =
\chi/3$, $C = 0$ and $a = 2$. In fact, we could have predicted that $C$ had
to vanish, because the radial coordinate in the post-TOV formalism is the
circumferential radius. With this in hand, the ppE parameters are then
simply
\begin{align}
\beta &=  \frac{5}{32}\, \chi \,  \eta^{-4/5} \,,
\qquad
\ppeb = -1 \,.
\label{eq:ppe_beta_and_b_final_mapping}
\end{align}
This is one of the main results of this paper, since a constraint on $\beta$ can now
straightforwardly be mapped to a constraint on $\chi$ and vice versa.
Note that one could also use the mapping between $(A,C,a) \to \chi$ to compute the
modification to Kepler's third law through Eq.~\eqref{eq:mod-kep} or the binding
energy as a function of the orbital frequency through
Eq.~\eqref{eq:bindng-e-of-w}, but this is not needed here.

In the limit $\chi = 0$ the evolution of a neutron star binary in GR and in
the post-TOV formalism become identical. However, we emphasize that this
limit \textit{does not necessarily} correspond to the limit in which the
post-TOV equation reduces to the usual GR TOV equations. Indeed, $\chi = 0$
only places a constraint on the combination of some of the post-TOV
parameters. Therefore, one can have the situation in which a neutron star
binary inspiral is identical to GR, yet the structure of the individual
stars is different from GR either because $\pi_2 - \mu_2 - 2 \pi \mu_1 = 0$
and/or because the nonzero post-TOV parameters are the ones which do not
affect the exterior space. Thus, we will refer to the case $\chi = 0$ as
the \textit{coincident limit}.

\section{Constraints on the post-TOV parameters from GW170817}
\label{sec:constraints}

The LVC released constraints on model-independent deviations from GR to
examine the consistency of the GW170817 event with GR
predictions~\cite{Abbott:2018lct,LVC:grtest}.
The constraints were obtained using a variant of
\textsc{IMRPhenomPv2}~\cite{Ajith:2007qp,Ajith:2009bn,Santamaria:2010yb,Husa:2015iqa},
which improves upon \textsc{IMRPhenomD}~\cite{Husa:2015iqa,Khan:2015jqa} by
phenomenologically including some aspects of spin precession
and tidal effects~\cite{Dietrich:2017aum,Dietrich:2018uni}.
In this variant, deviations from GR are described through relative shifts
in the GR PN coefficients of the Fourier phase of \textsc{IMRPhenomPv2}
\begin{equation}
    \phi_{i} \rightarrow \phi_{i} \left(1 + \delta
    \phi_{i}\right)\,,
\end{equation}
where $\delta \phi_{i}$ are additional free parameters in the model.

The parametrization used by LVC is an implementation of the ppE formalism
as explained in~\cite{Yunes:2016jcc}, with $\beta$ and $\delta\phi_{4}$
being related as
\begin{equation}
    \beta = \frac{3}{128} \, \phi_{4} \, \delta\phi_{4} \, \eta^{-4/5}\,,
    \label{eq:beta_ppe}
\end{equation}
where $\phi_{4}$ is the GR coefficient of the Fourier phase at 2PN order
(cf.~Appendix B in~\cite{Khan:2015jqa}).
Comparing Eqs.~\eqref{eq:ppe_beta_and_b_final_mapping} and~\eqref{eq:beta_ppe} we obtain
\begin{equation}
    \chi = \frac{3}{20} \, \phi_4 \, \delta \phi_4\,,
    \label{eq:chi_fun_deltaphi}
\end{equation}
which establishes the relation between $\delta \phi_{4}$ with $\chi$.

We can now translate the posterior distribution of $\delta \phi_4$ into one
for $\chi$ by using the MCMC samples available in~\cite{LVC:grtest}, where, for each
step, we calculate the corresponding value of $\chi$ using
Eq.~\eqref{eq:chi_fun_deltaphi}. The resulting probability density is shown
in Fig.~\ref{fig:chi_posterior} with the 90\% credible region corresponding
to
\begin{equation}
-2.4 \leq \chi \leq 44\,.
\label{eq:constraint_chi}
\end{equation}
This is the first constraint on (a combination of) post-TOV parameters and
another one of the main results of this paper.

The fact that the posterior of $\chi$ has a peak outside of zero (the
coincident limit) is perplexing at first sight and may be misinterpreted as
evidence for a deviation from GR, but \emph{this is not to be the case}.
Rather, it reflects the qualitative behavior of the posterior distribution
of $\delta \phi_4$~(see Fig. 1 in \cite{Abbott:2018lct}), which also does
not exhibit a peak at $\delta \phi_4 = 0$ and it is skewed to positive
values.  Both distributions, however, clearly do have a significant amount
of support at zero, and thus, they do not indicate an inconsistency with
GR. The skewness in the posterior for $\delta\phi_{4}$ probably results
from the marginalization process over the various parameters that describe
the model, the degeneracies between these parameters, and the
nonstationarity of the noise in the detectors.

The similarity between the posteriors for $\chi$ and $\delta \phi_4$ can be
understood from the following argument. The two posteriors, $P(\chi)$ and
$P(\delta \phi_4)$, are related by $P(\chi) = P(\delta
\phi_4) (\dd \delta \phi_4 / \dd \chi)$. The Jacobian of the transformation
($\dd \delta \phi_4 / \dd \chi$) can be calculated from
Eq.~\eqref{eq:chi_fun_deltaphi}, where $\phi_4$ is independent of $\delta
\phi_4$. From the MCMC samples we find that the mean value of the prefactor
is $(3/20) \times \phi_{4} \approx 12.6$ and thus $P(\chi) \approx  P(\delta\phi_4) / 12.6$.
Moreover, $\chi \approx 12.6\, \delta\phi_4$, which streches $P(\chi)$ relative
to $P(\delta\phi_4)$.
We then come to the conclusion that $P(\chi)$ is
nothing but a rescaled version (by the same scale factor) in height and width of $P(\delta
\phi_4)$. In fact, this simple argument results in a posterior for $\chi$ that is
very similar to that shown in Fig.~\ref{fig:chi_posterior}.

Having obtained a constraint on $\chi$, is it possible to translate it into
constraints on the three-dimensional parameter space spanned by $\mu_{1}$,
$\mu_{2}$ and $\pi_{2}$?
The first step to do this, is to fix the prior ranges for these
parameters.
We take $\mu_1 \in [-1.0, 0.1]$ (for the reasons discussed in
Sec.~\ref{sec:ptov}) and assume $\mu_2$ and $\pi_2$ are in the ranges $[-22,
22]$.
The latter domains are chosen such as to include the GR limit and to be large
enough to include moderately large values of $\mu_2$ and $\pi_2$ to encompass the
upper bound $\chi = 44$.
We then draw samples from the probability density function $P(\chi)$ (shown in
Fig.~\ref{fig:chi_posterior}), and given a value $\chi_{i}$, we then draw samples
of $\mu_{1}$, $\mu_{2}$ and $\pi_{2}$ until Eq.~\eqref{eq:chi_def} is satisfied.

\begin{figure}[t]
    \includegraphics[width=0.95\columnwidth]{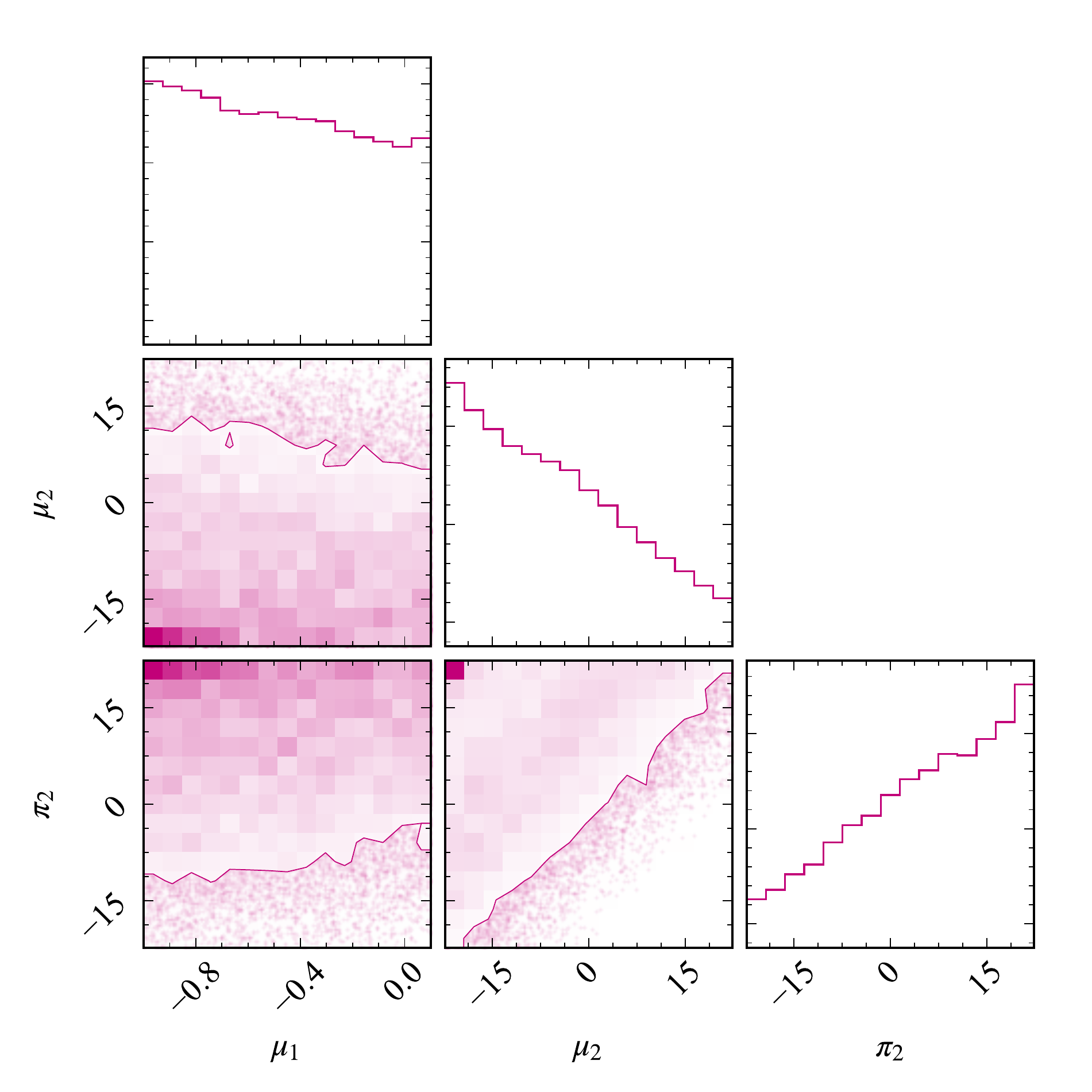}
\caption{Corner plot showing the posterior for $(\mu_1,\, \mu_2,\, \pi_2)$, as well as the
allowed 90\% credible regions (solid contour lines in the off-diagonal panels).
We see that whereas $\mu_1$ is essentially unconstrained, values of $\mu_2$
($\pi_2$) which are smaller (larger) are favored with peaks located at the
boundary of our prior ranges.
The strong degeneracy between these parameters follows from the fact
that the constraints derive from an underconstrained system, only requiring to satisfy
Eq.~\eqref{eq:chi_def}.
}
\label{fig:parameter_space}
\end{figure}

Figure~\ref{fig:parameter_space} shows the result of this calculation. The
diagonal panels in this corner plot show the marginalized posteriors on
$\mu_{1}$, $\mu_{2}$ and $\pi_{2}$, while the off-diagonal panels show
two-dimensional joint posteriors with the 90\% credible contours delimited
by the solid lines. The constraint on $\chi$
leaves $\mu_{1}$ essentially unconstrained, while the favored values
for $\mu_2$ and $\pi_2$ are set by the bound of our priors.
This occurs due to the strong degeneracy between these parameters arising
from Eq.~\eqref{eq:chi_def}, which, together with Eq.~\eqref{eq:constraint_chi}, constrains $
\pi_2 - (\mu_2 + 4 \pi \mu_1) < \textrm{const}$.
Thus, if the prior ranges of $\mu_2$ and $\pi_2$ were
extended, the marginalized posteriors in Fig.~\ref{fig:parameter_space}
would retain their qualitative shapes, with peaks at the edge of their priors,
as $\pi_2 - \mu_2 = \textrm{const}$ has an infinite number of solutions.

\section{Degeneracies between matter and gravity models}
\label{sec:mr}

In the previous section we have constrained the magnitude of the post-TOV
parameter $\chi$, as well as $\mu_1$, $\mu_2$ and $\pi_2$.
How do these results impact the allowed deformations away from a GR
mass-radius curve as allowed by the post-TOV formalism?
Could one, for example, use these deformed mass-radius regions, together with
observations of the mass and radius of isolated neutron stars, to place further
constraints on post-TOV parameters?
We will show in this section explicitly that this is not possible due to degeneracies
between post-TOV deformations and the EOS.

To answer this question, we construct mass-radius curves with a \textit{restricted} set of post-TOV
equations and a fixed set of representative EOSs. The set of post-TOV equations  is obtained from Eq.~\eqref{eq:postTOV}
by fixing all parameters to zero other than $\mu_1$, $\mu_2$ and $\pi_2$, and we make this choice
because these three parameters are the only ones that can be directly probed by
electromagnetic or gravitational wave phenomena. The set of EOSs consists of the
SLy~\cite{Douchin:2001sv} and APRb~\cite{Akmal:1998cf} EOSs, which are
favored by the tidal deformability measurements of the constituents of
GW170817~\cite{Abbott:2018exr} in GR and the observation of two
solar masses neutron stars~\cite{Demorest:2010bx,Antoniadis:2013pzd,Cromartie:2019kug}.
With this set of post-TOV equations and EOSs, we then construct one thousand mass-radius curves
each with a different choice of post-TOV parameters that lay within the bound of Eq.~\eqref{eq:constraint_chi}.
The value of these parameters was selected as follows.
First, we drew random samples from the probability distribution
function $P(\chi)$, only accepting values that satisfy~\eqref{eq:constraint_chi}.
Next, we drew samples of $\mu_1$, $\mu_2$ and $\pi_2$ (as in
Sec.~\ref{sec:constraints}) until Eq.~\eqref{eq:chi_def} is met.

\begin{figure}[t]
    \includegraphics[width=\columnwidth]{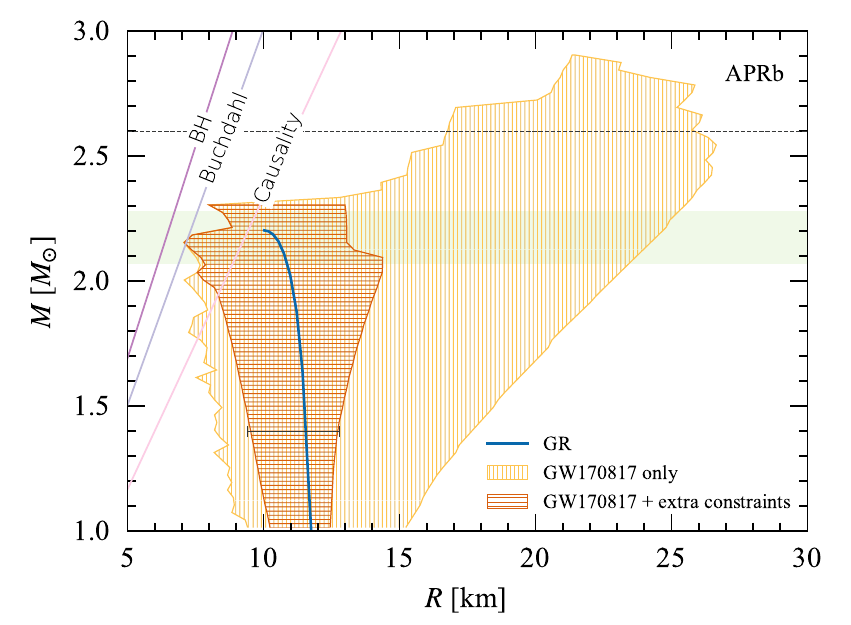}
\caption{Allowed modifications to the mass-radius relation of neutron stars
under the constraint $-2.4 \leq \chi \leq 44$, for EOS
APRb~\cite{Akmal:1998cf} (the case for SLy is qualitatively similar).
The vertically hatched regions represent the allowed post-TOV deformations
to GR the GW170817 constraint on $\chi$ only, while the solid line represents the GR result.
Requiring that additional constraints be satisfied, such as the mass
measurement of MSP J0740+6620~\cite{Cromartie:2019kug} ($M =
2.17^{+0.11}_{-0.10}$ M\msun, shaded region) and the radius of canonical
neutron stars~\cite{Kumar:2019xgp} ($R_{1.4} = 10.9^{+1.9}_{-1.5}$ km,
horizontal solid line) the allowed region is reduced to the horizontally
hatched region.
For reference, we also included in the limit set by Schwarzschild BHs ($R=2M$),
Buchdahl's limit ($R = 9M/8$), the limit set by causality ($R = 2.9
M$)~\cite{Rhoades:1974fn,Kalogera:1996ci} in GR and the cut-off mass
(dotted line) $M = 2.6$ M\msun\, inferred from the mass distribution of compact
binaries containing neutron stars~\cite{Alsing:2017bbc}.
}

\label{fig:mass_radius_details}
\end{figure}

\begin{figure}[t]
    \includegraphics[width=\columnwidth]{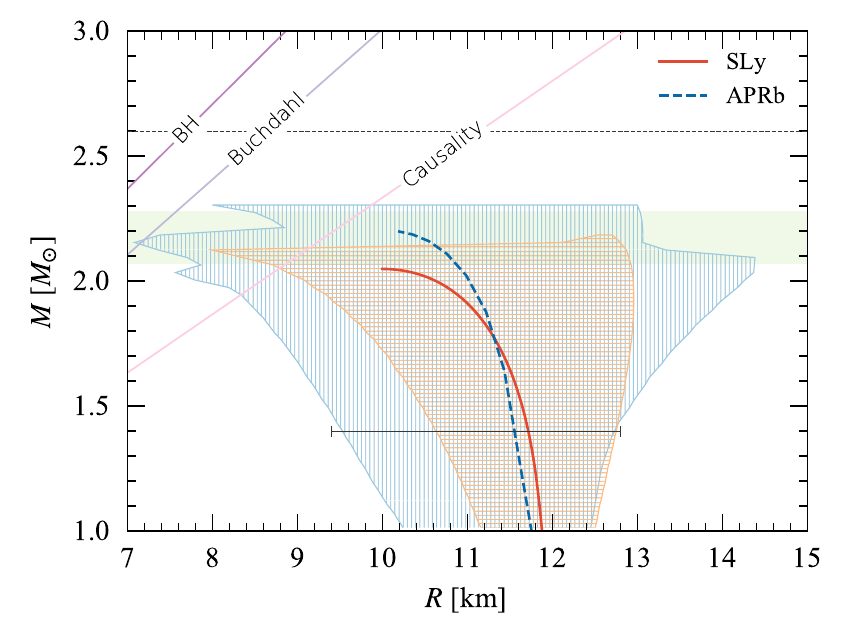}
\caption{Similar to Fig.~\ref{fig:mass_radius_details}, however
only showing the more restrictive regions for both EOSs, APRb and SLy.
This figure explicitly shows the degeneracies between EOSs assuming a
theory of gravity to be known (see the solid and dashed curves) and theory of
gravity assuming that the EOS is known \textit{a priori} (individual hatched
regions).
Varying both EOS and theory of gravity increases further the degeneracy
between matter and gravity models - a degeneracy due to the fact
that neutron stars are relativistic objects.
}
\label{fig:mass_radius_main}
\end{figure}

The results of these integrations are shown in
Fig.~\ref{fig:mass_radius_details} for EOS APRb; the results for EOS
SLy being very similar, so we do not show them here. In this figure, the vertical hatched
(yellow) region contains all the mass-radius curves that are consistent with the post-TOV
constraints derived in this paper, all truncated at the the maximum mass of
the (stable) sequence.
As is evident, the post-TOV formalism is capable of capturing a wide variety
of curves that span a large region of the mass-radius plane, including
exotic types, which e.g. have very low maximum masses $M_{\rm max}
\approx 1.5$ M\msun (despite both EOSs supporting $\gtrsim 2$ M\msun\,
stars in GR). Other curves can enter the region in the mass-radius plane that
is excluded in GR (the ``causality'' curve), which is derived by requiring only
a very minimal set of assumptions on the underlying unknown EOS~\cite{Rhoades:1974fn,Kalogera:1996ci}, with
some even extending close to Buchdahl's limit.\footnote{This
high-compactness stars are supported in the post-TOV formalism due to
the fact that the $\calp_{2}$ modification can be associated to
\textit{pressure anisotropy}, with the difference between radial and
tangential pressures being $p_{r} - p_{t} = \rho\, m\,\calp_{2} / (2r)$
[see Eq.~\eqref{eq:second_set}]. Pressure anisotropy has have long been
known to support ultracompact stars~\cite{1974ApJ...188..657B}.}
Further exotica include mass-radius curves that do not
have an extrema at $M_{\rm max}$. These generically allow for very large
radii ($\gtrsim 15$ km), even when the mass is $1.4$ M\msun\,.
More common curves are only small deformations away from the GR result.

Although the region of the mass-radius plane
allowed by Eq.~\eqref{eq:constraint_chi} \textit{alone} is rather large,
it can be reduced by combining other sources of
information on the masses and radii of neutron stars.  For instance, by
imposing that the mass-radius curves are consistent with (i) the existence
of neutron stars with masses $M = 2.17^{+0.11}_{-0.10}$
M\msun\,~\cite{Cromartie:2019kug} and (ii) the canonical radius bound
$R_{1.4} = 10.9^{+1.9}_{-1.5}$ km~\cite{Kumar:2019xgp}, then 99.3\% (for
SLy) and 96.3\% (for APRb) of the curves investigated are excluded. The
resulting tighter contour due to the surviving mass-radius curves is shown
by the horizontally hatched (red) regions in Fig.~\ref{fig:mass_radius_details}.

Figure~\ref{fig:mass_radius_main} vividly shows several difficulties in testing
extreme gravity with observations of isolated neutron stars that yield
mass and radius measurements alone.
First, even in GR, our ignorance on the underlying neutron star EOS gives
riseto mass-radius curves that can overlap (see the intersection of the SLy
and APRb curves in Fig.~\ref{fig:mass_radius_main}).
Second, even in the event of the EOS being tightly constrained in the future
(under the assumption of neutron stars are described by GR), a measurement
of $\chi$ still leads to degeneracies between the post-TOV parameters
$\mu_{1}$, $\mu_{2}$ and $\pi_{1}$, as shown in the previous section
Each value of $(\mu_{1},\mu_{2},\pi_{1})$ should correspond to a specific theory
of gravity, and this degeneracy prevents us from singling one out.
Third, the fact that the contours in Fig.~\ref{fig:mass_radius_main} change as
we change the EOS makes the degeneracy between EOS and theory
of gravity explicit.
This degeneracy arises in the post-TOV formalism in a very explicit way: the post-TOV equations (with
$\calp_1 = \calm_1 = 0$) can be mapped into an effective barotropic EOS, with
$p = p (\varepsilon_{\rm eff})$ and $\varepsilon_{\rm eff} \equiv
\varepsilon + \rho \calM_2$~\cite{Glampedakis:2015sua}.
Therefore, observations of isolated neutron stars that yield
mass and radius measurements alone cannot really be used to test gravity,
unless more information is contained in the data, which can be folded into the models
to test GR.

\section{Conclusions and outlook}
\label{sec:conclusions}

Neutron star observations, both through electromagnetic and gravitational-wave astronomy,
offer us a unique look into the fundamental interactions of nature.
For gravity (in particular) it allows us to probe both the strong-field regime of
neutron star interiors and the radiative aspects of gravity, when these object
are found in binary systems.
To be able to do theory-independent tests of gravity through neutron star
observations, we have combined the post-TOV and ppE formalisms,
constructing a
single, unified framework for which tests of gravity can be performed from the
radiative level down to the level of stellar structure.

This framework is particularly relevant in light of ongoing events on the
observational front. For instance, the \textit{Neutron Star Interior
Composition Explorer} (NICER)
mission~\cite{2012SPIE.8443E..13G,2014SPIE.9144E..20A,2017NatAs...1..895G}
will soon release the mass and radius measurements of a number of neutron
stars within 10\% precision and probes the effects of spacetime curvature
on the motion of photons.
Moreover, LIGO/Virgo is currently on its
third scientific observing run, with a binary neutron star merger candidate
already observed and tens of events expected to be seen in the next
years.
It would be interesting to combine these upcoming observational results to further
explore the resulting constraints on the post-TOV parameters and thereby
constrain modifications to GR in a theory-independent way.

For instance, the contours in Fig.~\ref{fig:mass_radius_main} reveal that the
largest variability occurs for massive stars with $M \gtrsim 1.8$ M\msun. One
of NICER's targets (PSR J1614–2230) has a mass of 1.93
M$_{\odot}$~\cite{Fonseca:2016tux,Miller:2016kae} and a radius measurement of
it would constrain this region of the mass-radius plane.
In turn, these constraints could also be used to probe deviations from GR in a
number of astrophysical scenarios, for instance in the quasiperiodic
oscillations on matter disks in accreting neutron
stars~\cite{Glampedakis:2016pes}, or in the pulse profiles emitted by hot spots on
the surface of rotating neutron stars (complementary to constraints on
scalar-tensor gravity~\cite{Silva:2019fqs}).
We have here only taken a first step on using this new framework and hope to
explore further its applications in the near future.

\section*{Acknowledgments}
We thank the post-TOV practitioners Emanuele Berti, Kostas Glampedakis and
George Pappas for numerous discussions on the topic over the years.
We also thank Alejandro C\'ardenas-Avenda\~{n}o, Katerina Chatziioannou, Remya
Nair, Thomas Sotiriou and Jacob Stanton for discussions on different aspects related to
this work.
Finally, we thank the anonymous referee for carefully reading our work.
This work was supported by NASA Grants No.~NNX16AB98G and No.~80NSSC17M0041.
N.~Y. also acknowledges the hospitality of KITP where some of this work was completed.

\appendix
\section{Derivation of the binding energy}
\label{app:binding}

In this Appendix, we derive general formulas for the changes to the energy
and angular momentum of point particles orbiting in the static, spherically
symmetric spacetime of an object of mass $M$.

\subsection{Particle motion in perturbed spacetimes}

Consider the line element
\begin{equation}
    \dd s^2 = - f(r) \dd t^2 + h(r) \dd r^2 + r^2 (\dd\theta^2 + \sin^2\theta \,\dd\phi^2)\,,
    \label{eq:line_element}
\end{equation}
in Schwarzschild coordinates, on which a massive particle follows geodesic
motion, with trajectory $x^{\alpha}(\tau)$, where
$\tau$ is the proper time. Let $u^{\alpha} \equiv \dd x^{\alpha} / \dd \tau$ be the
particle's four-velocity, constrained by $g_{\alpha\beta} u^{\alpha} u^{\beta} = - 1$

As usual, the spacetime symmetries imply the existence of two Killing vector fields which
result in two conserved quantities
\begin{equation}
    E \equiv - g_{tt} \dot{t}\,, \quad
    L \equiv g_{\phi\phi} \dot{\phi}\,,
    \label{eq:conserved}
\end{equation}
respectively, the energy and angular momentum (per unit mass) of the particle.

Due to the conserved angular momentum, orbits are confined to a single plane, which we
take, without loss of generality to be the one for which $\theta = \pi / 2$.
Using this result we find that
\begin{equation}
    \frac{1}{2} E^2 = \frac{1}{2} f(r) h(r) \dot{r}^2 + \frac{1}{2} f(r) \left(\frac{L^2}{r^2} + 1\right)\,.
    \label{eq:energy_balance}
\end{equation}

Let us consider spacetimes with metric $\bg$, which are a small deformations to
a static, spherically symmetric background $\bg_0$.
More specifically, let us write the metric functions $f$ and $h$ as
\begin{equation}
    f(r) \equiv f_0(r) + \varepsilon f_1(r)\,, \quad
    h(r) \equiv f^{-1}_0(r) + \varepsilon h_1(r)\,,
\end{equation}
where (in this Appendix only) $\varepsilon$ denotes a small bookkeeping parameter.
For convenience, we omit hereafter the dependence on $r$ of the functions
introduced above.
Using these decompositions of $f$ and $h$, into Eq.~\eqref{eq:energy_balance} and
then solving for $\dot{r}^2$, we find to leading order in $\varepsilon$,
\begin{align}
    \frac{1}{2}\dot{r}^2 &= \frac{1}{2}E^2 - \frac{1}{2} f_0 \left( \frac{L^2}{r^2} + 1\right)
    \nonumber \\
    &\quad -\frac{1}{2} \varepsilon f_0 h_1 \left[ E^2 - f_0 \left(  \frac{L^2}{r^2} + 1 \right)\right]
    - \frac{1}{2} \varepsilon E^2 f_1 f_0^{-1}\,.
    \nonumber \\
    \label{eq:perturbed_energy_balance}
\end{align}
Equation~\eqref{eq:perturbed_energy_balance} suggests the definition of a zeroth-order
effective potential $V^0_{\rm eff}$,
\begin{equation}
    V^0_{\rm eff} \equiv \frac{1}{2} E^2 - \frac{1}{2} f_0 \left( \frac{L^2}{r^2} + 1 \right)\,,
\end{equation}
and a leading-order correction $V^{1}_{\rm eff}$,
\begin{equation}
    V^{1}_{\rm eff} \equiv -\frac{1}{2} E^2 f_1 f_0^{-1} - f_0 h_1 V^0_{\rm eff}\,,
\end{equation}
such that Eq.~\eqref{eq:perturbed_energy_balance} becomes
\begin{equation}
    \frac{1}{2}\dot{r} = V^0_{\rm eff} + \varepsilon V^{1}_{\rm eff}\,.
\end{equation}

\subsection{Properties of particles in circular orbits}

Now let us focus on the properties of particles in (not necessarily stable) circular orbits
that we denote by $\rc$.
These orbits satisfy the conditions
\begin{equation}
    \dot{r} = 0\,, \qquad \dd V_{\rm eff} / \dd r = 0\,,
    \label{eq:circular_conditions}
\end{equation}
where $V_{\rm eff} \equiv V^{0}_{\rm eff} + \varepsilon V^{1}_{\rm eff}$.

As a warm-up exercise, let us consider the limit $\varepsilon \to 0$ and obtain
general formulas of the (zeroth-order) energy $E_0$ and angular momentum $L_0$
of particles in circular orbits on $\bg_0$.
This calculation is particularly simple, because $L_0$ can
be easily isolated from the $\dd V_{\rm eff} / \dd r$ equation. With a
little algebra we can obtain the general formulas
\begin{align}
    L_0^{2} &= - \left. \left(\frac{\dd f_0}{\dd r} \right)
    \left[ \frac{\dd (f_0 r^{-2})}{\dd r} \right]^{2}\right\vert_{r = \rc} \,.
    \label{eq:l_zero}
    \\
    E_0^2 &= \left. f_0(\rc) \left\{
    1 - \frac{1}{r^2}\left[ \frac{\dd (f_0 r^{-2})}{\dd r} \right]^{2}
    \right\} \right\vert_{r = \rc}\,.
    \label{eq:e_zero}
\end{align}
%
%
In the particular limit of the Schwarzschild spacetime ($f_0 = 1 - 2 M /
r$) we readily obtain the familiar results
\begin{equation}
    L^2_0 = \frac{M \rc}{1 - 3 M / \rc}\,, \quad
    E^2_0 = \frac{(1 - 2 M / \rc)^2}{1 - 3 M / \rc}\,.
\end{equation}

Now, let us consider the general problem and obtain the corrections to $E_0$ and $L_0$ due to the
perturbation $V^{1}_{\rm eff}$ in Eq.~\eqref{eq:perturbed_energy_balance}.
To do this, we first solve Eq.~\eqref{eq:circular_conditions} for $E^{2}$ and $L^{2}$. Next, we expand the
resulting expressions to leading order in $\varepsilon$.
The outcome of this exercise is that $E$ and $L$ can be written as
\begin{equation}
    E^2 = E_0^{2} + \varepsilon E_1^{2} \,, \quad
    L^2 = L_0^{2} + \varepsilon L_1^{2} \,,
\end{equation}
where the corrections to the zeroth-order energy and angular momentum [cf. Eqs.~\eqref{eq:l_zero}
and~\eqref{eq:e_zero}] are
\begin{align}
    E^2_1 &= \frac{2 f_0}{r^6}
            \left(
            2 f_0 f_1 - 2 r f_1 \frac{\dd f_0}{\dd r} + r f_0 \frac{\dd f_1}{\dd r}
            \right)
            \nonumber \\
            &\quad  \times \left. \left[ \frac{\dd (f_0 r^{-2})}{\dd r} \right]^{-2}
            \right \vert_{r = \rc}\,,
    \label{eq:e_correction}
\end{align}
and
\begin{equation}
    L^2_1 = \left. -\frac{2}{r^3} \frac{\dd (f_0 f_1)}{\dd r}
    \left[ \frac{\dd (f_0 r^{-2})}{\dd r} \right]^{-2} \right \vert_{r = \rc}\,.
    \label{eq:l_correction}
\end{equation}
These expressions are the main result of this Appendix. Notice
the absence of $h_1$ in these expressions.

Finally, we can solve for $E$ and $L$ and write
\begin{align}
    E &= E_0 + \delta E \equiv E_0 + \varepsilon \frac{E_1^2}{2 E_0}\,,
    \label{eq:e_final} \\
    L &= L_0 + \delta L \equiv L_0 + \varepsilon \frac{L_1^2}{2 L_0}\,,
    \label{eq:l_final}
\end{align}
our final results.

We emphasize that although the formulas obtained here were applied for the
post-Schwarzschild metric, our results can be used to \textit{any} perturbed
spacetime - as long as its line element can be written in the form
of~\eqref{eq:line_element} - and then connected to the
ppE formalism through Eq.~\eqref{eq:ppe_beta_and_b}.

\section{Orbital period decay rate}
\label{app:pdot}

In this appendix we derive an expression for the orbital period
rate of change $\dot{P}$ in the post-TOV formalism following
closely~\cite{Sampson:2013wia} and obtain an order-of-magnitude
bound on $\chi$ from binary systems.

We start by assuming that energy is carried away from a circular
binary according to the GR gravitational-wave luminosity formula
\begin{equation}
    \dot{E} = \frac{32}{5}\, \eta^2\, \frac{m^5}{r^5}\,,
    \label{eq:gw_lum}
\end{equation}
at the expense of the orbital binding energy given
by~\eqref{eq:bindng-e-of-w}, i.e. $\dot{E}_{\rm b} = -\dot{E}$.

Taking a time-derivative of Eq.~\eqref{eq:bindng-e-of-w} and
using $\omega = 2 \pi / P(t)$ we find:
\begin{align}
    \dot{E}_{\rm b} &= - \frac{1}{3} \mu \left(\frac{2 \pi m}{P} \right)^{2/3}
    \, \frac{\dot{P}}{P}
    \nonumber \\
     &\quad \times \left[1
    - \frac{2}{3}A(2a-1)(a+1)\left( \frac{2 \pi m}{P} \right)^{2a/3}
    \right.
    \nonumber \\
    & \quad\left.+ \frac{1}{3}C(2c-1)(c+1) \left( \frac{2 \pi m}{P} \right)^{2c/3}
    \right]\,.
\label{eq:binding-e-dot}
\end{align}

Now, let us return to~\eqref{eq:gw_lum}. We can eliminate $r$ in
favor of $\omega$ by using the modified Kepler's
law~\eqref{eq:mod-kep}. Solving for $r$, expanding in $A$, $C$ and
then substituting the resulting expression in Eq.~\eqref{eq:gw_lum}
gives
\begin{align}
    \dot{E} &= \frac{32}{5} \eta^2 \left(\frac{2 \pi m}{P} \right)^{10/3}
    \left\{
        1 - \frac{5}{3} \left[ A(a+1) \left( \frac{2 \pi m}{P} \right)^{2a/3} \right.
    \right.
    \nonumber \\
            &\quad - \left. \left. \frac{1}{2}
            C (c-2) \left( \frac{2 \pi m}{P} \right)^{2c/3}
            \right] \right\} \,.
    \label{eq:gw_lum_p}
\end{align}
We can now use Eqs.~\eqref{eq:binding-e-dot}
and~\eqref{eq:gw_lum_p} in the energy balance law, solve for
$\dot{P}$ (while expanding once more in $A$, $C$) and find:
\begin{align}
    \frac{\dot{P}}{P} &= \left(\frac{\dot{P}}{P}\right)_{\rm GR}
    \left[1+\frac{1}{3} A (4a-7) (a+1)
        \left(\frac{2\pi m}{P}\right)^{2a/3}
    \right.
    \nonumber \\
    &\quad \left.
        - \frac{1}{6} C (4 c^2 -3c + 8) \left(\frac{2\pi m}{P}\right)^{2c/3}
    \right]\,,
\end{align}
which is the main result of this appendix, where
\begin{equation}
    \left(\frac{\dot{P}}{P}\right)_{\rm GR} = - \frac{96}{5} \frac{\eta^2}{\mu}
                                            \left( \frac{2\pi m}{P} \right)^{8/3} \,,
\end{equation}
is the corresponding GR result.
In the particular case of the post-TOV metric, we find after
using $A = \chi / 3$, $C=0$ and $a=2$ that
\begin{equation}
    \frac{\dot{P}}{P} = \left( \frac{\dot{P}}{P}\right)_{\rm GR}
    \left[1 + \frac{1}{3}\,\chi\,\left(\frac{2 \pi m}{P}\right)^{4/3}\right] \,.
    \label{eq:pdot_ptov}
\end{equation}

A simple constraint on $\chi$ (independent from the one in the main
text) can thus be obtained as follows. Since binary pulsar
observations of $(\dot{P}/P)_{\rm obs}$ are in remarkable agreement
with GR up to some observational error $\delta$ we can write
$(\dot{P}/P)_{\rm obs}=(\dot{P}/P)_{\rm GR}(1+\delta)$.
Therefore, the post-TOV correction in Eq.~\eqref{eq:pdot_ptov} is
bound by $\delta$, which then constrains $\chi$ to be
\begin{equation}
    |\chi| \leq 3 \,\delta \left(\frac{P}{2 \pi m}\right)^{4/3} \approx \delta \, v_{\rm c}^{-4}\,,
\end{equation}
where $v_{\rm c} \approx 2.1 \times 10^{-3}$ is characteristic velocity of the
system and $\delta \approx 1.3 \times 10^{-2}$ (for
the quasicircular system PSR J0737-3039~\cite{Yunes:2010qb}), giving the weak bound $|\chi|
\lesssim 7.2 \times 10^{8}$. This result is seven orders of magnitude weaker than
the bound obtained from GW170817 and exemplifies the constraining power
of gravitational wave events on modifications to GR relative to binary
pulsar constraints\footnote{At first sight our upper
bounds on $|\chi|$ are outside the perturbative
regime ($A,B,C \ll 1$) used to derive our main formulas. These
expansions are only formal. The \emph{true small parameter} bound
to be $\ll 1$ is the combination, e.g. $A v^{a}$
(similarly for the other parameters) which does remain small during the
inspiral.}.

\bibliographystyle{apsrev4-1}
\bibliography{biblio}

\end{document}